\documentclass[%
 reprint,
 amsmath,amssymb,
 aps,
]{revtex4-2}

\usepackage{graphicx}
\usepackage{dcolumn}
\usepackage{bm}
\usepackage{subfigure}
\usepackage{gensymb}
\usepackage{natbib}
\usepackage{color}
\usepackage{textcomp}

\usepackage[colorlinks = true,
            linkcolor = blue,
            urlcolor  = blue,
            citecolor = blue,
            anchorcolor = blue]{hyperref}
\usepackage{etoolbox}
\usepackage{mathtools}
\usepackage{siunitx}

\begin{document}

\preprint{APS/123-QED}

\title{Spontaneous rolling on a 90$^{\circ}$ incline}

\author{Surjyasish Mitra$^{1}$, A-Reum Kim$^{2}$, Boxin Zhao$^{2}$, and Sushanta K. Mitra$^{1,\ast\ast}$\\
\vspace{0.5 cm}
\normalsize{$^{1}$ Department of Mechanical and Mechantronics Engineering,}
\normalsize{Waterloo Institute for Nanotechnology, University of Waterloo,}
\normalsize {200 University Avenue West, Waterloo, ON N2L 3G1, Canada}\\
\normalsize{$^{2}$ Department of Chemical Engineering, Waterloo Institute for Nanotechnology, University of Waterloo,}
\vspace{0.5 cm}
\normalsize{$^{\ast\ast}$Corresponding author: 
skmitra@uwaterloo.ca}}


%
%


\date{\today}

\begin{abstract}
On perfectly vertical surfaces, rolling is conventionally deemed impossible without external torque. While various species like geckos and spiders exhibit vertical locomotion, they cannot achieve rolling; instead, they fall. In this study, we demonstrate the spontaneous rolling of an elastic polyacrylamide sphere on an elastic polydimethylsiloxane (PDMS) substrate held vertically at a 90$^{\circ}$ incline, given specific elasticity values for the materials. Our experiments uncover a slow rolling motion induced by a dynamically changing contact diameter and a unique contact asymmetry. The advancing edge behaves like a closing crack, while the receding edge acts as an opening crack. Utilizing adhesion hysteresis theories and crack propagation models, we explain how this contact asymmetry generates the necessary torque and friction to maintain rolling, preventing either pinning or falling. The findings challenge the traditional understanding of vertical surface interactions and open new avenues for exploring soft-on-soft contact systems. This novel phenomenon has potential implications for designing advanced materials and understanding biological locomotion on vertical surfaces.
\end{abstract}

\maketitle

From automobile wheels to the game of billiards, rolling is ubiquitous. Typically, we observe rolling on horizontal or inclined surfaces. Revisiting introductory physics textbooks \cite{halliday2013fundamentals}: when a rigid glass sphere, for instance, of mass $m$ is gently dispensed on an inclined plane of a similar rigid material, the sphere will roll down freely due to gravity for any inclination angle $0^{\circ} < \theta_{\rm S} < 90^{\circ}$. But what happens when $\theta_{\rm S} = 90^{\circ}$, i.e., a perfectly vertical surface? The intuitive answer is that the sphere will undergo a free fall without any contact with the surface unless it has an initial roll (see movie S1). This is because, on a perfectly smooth vertical surface, the normal force at the contact interface (or point), i.e., $mg\cos{90^{\circ}}$ is zero, resulting in no static friction force to generate the torque needed for rolling. Typically, rigid-on-rigid contacts have little to no adhesion \cite{hertz1882beruhrung}. For scenarios, where adhesion is significant \cite{johnson1971surface}, the rigid sphere will either stick to the surface or slide down. 

On the other extreme, a millimetric liquid water droplet will wet a perfectly vertical surface, forming a downward elongated profile and exhibiting finite contact angle hysteresis, which is the difference in contact angles between the advancing and receding edges (see movie S2). The droplet may either remain pinned in its original position or slide down freely while maintaining contact with the surface (see movie S2). Under special circumstances, for instance, on an inclined micro-nano textured superhydrophobic substrate, droplets may display irregular rolling and tumbling motion \cite{mukherjee2019surface}. However, rolling of any known object on a perfectly smooth, vertical surface has never been reported. The question is whether this overarching notion of inability of  objects to roll on surfaces when $\theta_{\rm S} = 90^{\circ}$ still holds true when both objects and underlying surfaces are made of elastic material (non-rigid) whose elasticity can be tuned at will? In this work, while exploring soft-on-soft contact/wetting systems across a broad range of both top and bottom contacting pair elasticity, we report a scenario where an elastic sphere exhibits spontaneous rolling motion on an elastic substrate held vertically at a 90$^{\circ}$ incline without any externally imparted torque. In the remainder of the text, we unravel the governing principles of our experimental observations by invoking a theory based on crack propagation and adhesion hysteresis.

We conduct our experiments by gently depositing 1\,mm radius polyacrylamide (PAAm) elastic spheres (mass, $m$ = 4 mg) on substrates kept vertically at 90$^{\circ}$ inclination with precise control using the software-controlled rotating stage of a commercial goniometer (DSA30, Kruss). The elastic spheres are realized using an in-house recipe involving a delicate mix of acrylamide (monomer),  N, N'-Methylene-bis-acrylamide (BIS) (crosslinker), and 2,4,6-tri-methyl benzoyldiphenylphosphine oxide (TPO) (initiator) \cite{kim2024flexible} (Fig.~S1, \emph{Supplementary Information} \cite{SI}). 
By tuning the monomer in weight percentages of 4 to 30, we create elastic spheres with elasticity, i.e., Young's modulus $E_{1}$ varying between 0.0017\,kPa to 169.7\,kPa (Fig.~S2 \citep{SI}). As our working substrates, we use freshly cleaned microscope glass slides and glass substrates coated with two different commercially available PDMS: Sylgard 184 and Sylgard 527 with monomer:crosslinker ratios of 10:1 and 1:1, respectively. Consequently, we prepare 1\,mm thick soft coatings with Young modulus, $E_{2} = 2242\,\rm{kPa}$ and  $E_{2} = 3\,\rm{kPa}$ (Fig.~S3 \citep{SI}). By varying the elasticity of both the elastic spheres and the substrates, we vary the effective elasticity of the contact interface, $E^{*} = [(1-{\nu_{1}}^2)/E_{1} + (1-{\nu_{2}}^2)/E_{2}]^{-1}$, where $\nu_{1}\approx 0.5$ and $\nu_{2}\approx0.5$ are the Poisson's ratio of the top and bottom pair, respectively  \cite{kim2024flexible}. 

For weakly elastic PAAm spheres, i.e., $E_{1} = 0.0017\,{\rm kPa}$, we observe that they behave as polymeric liquids and completely wet the substrates like a water droplet \cite{mitra2024rapid}. Consequently, for the substrates held vertically, they mostly remain pinned (Fig.~S4 \citep{SI}). This behavior is similar to water droplets which mostly remain pinned on the soft, PDMS substrates (Fig.~S5 \citep{SI}) whereas they slide down on glass substrates (see movie S2). The outcome depends on the substrate wettability and the presence of local surface heterogeneities \cite{joanny1984model} or contact line induced local deformation (for the soft PDMS coatings) \cite{jerison2011deformation,mitra2022probing, vo2024unsteady} which favors pinning rather than sliding. Since the focus of the present work is more on rolling, we do not elaborate further on the pinned or sliding outcomes. With increasing PAAm elasticity, we observe morphologies intermediate between wetting and contact with gradually decreasing PAAm-substrate contact diameter $d$ (Figs.~S6-S9 \citep{SI}). However, they still remain either pinned or exhibit sliding behavior. For $E_{1} = 15.8\,{\rm kPa}$, the elastic spheres exhibit the Hertz/Johnson-Kendall-Roberts (JKR) like contact configuration \cite{hertz1882beruhrung,johnson1971surface}; though still pinned to the substrates held vertically (Fig.~S7 \citep{SI}). For the stiffest PAAm sphere, i.e.,  $E_{1} = 169.7\,{\rm kPa}$, we observe an interesting outcome: the spheres exhibit \emph{steady rolling} (see movies S3-S5) on the relatively stiffer PDMS substrate ($E_{2} = 2242\,{\rm kPa}$) held vertically while remaining pinned on the glass and the relatively softer PDMS substrate ($E_{2} = 3\,{\rm kPa}$) (Fig.~S8 \citep{SI}).

\begin{figure}[b]
\begin{center}
\includegraphics[width=0.48\textwidth]{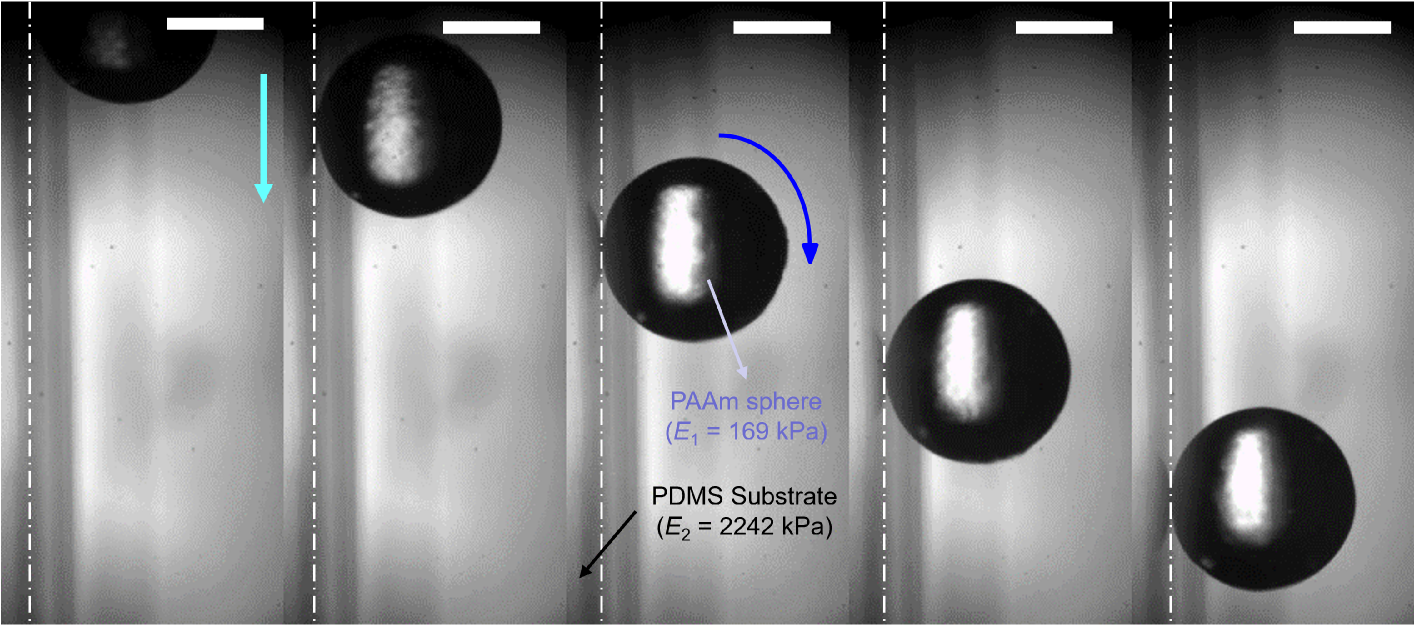}
\caption{Experimental snapshots highlighting the rolling of a 1\,mm radius PAAm sphere (elasticity, $E_{1} = 169.7\,{\rm kPa}$) on a vertical PDMS substrate (elasticity, $E_{2} = 2242\,{\rm kPa}$). The frames are separated by 2 seconds. The scale bars represent 1\,mm.}
\label{fig:1}
\end{center}
\end{figure}

\begin{figure*}[]
\begin{center}
\includegraphics[width=0.98\textwidth]{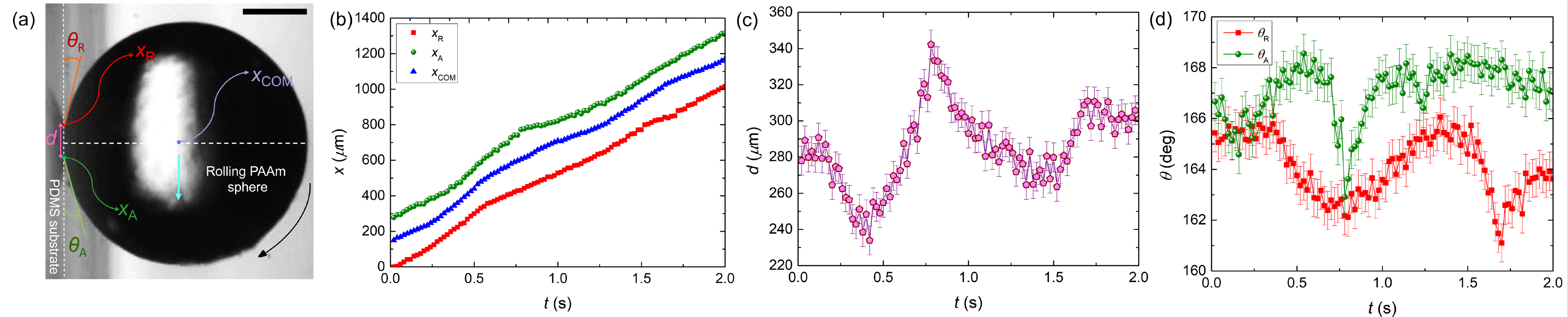}
\caption{(a) Experimental snapshot during rolling of a 1 mm radius PAAm sphere ($E_{1} = 169.7\,{\rm kPa}$) on a vertical PDMS substrate ($E_{2} = 2242\,{\rm kPa}$). The location of the advancing contact edge, receding contact edge, and the center-of-mass are highlighted as $x_{\rm A}$, $x_{\rm R}$, and $x_{\rm COM}$, respectively. $\theta_{\rm A}$ and $\theta_{\rm R}$ are the contact angles at the advancing and receding edge, respectively. $d$ is the contact diameter. The arrow denotes the direction of motion. The scale bar represents 0.5\,mm.  (b) Evolution of the advancing edge $x_{\rm A}$, receding edge $x_{\rm R}$, and the center-of-mass $x_{\rm COM}$. Note that the location of the receding edge at the first frame is considered the origin, i.e., $x= 0$. Error bars are avoided for clarity. Measurement error is 7-8\,$\mu$m. (c) Evolution of the contact diameter $d$. (d) Evolution of the contact angles at the advancing edge $\theta_{\rm A}$ and the receding edge $\theta_{\rm R}$.}
\label{fig:2}
\end{center}
\end{figure*}

Fig.~\ref{fig:1} shows the snapshots of the rolling event (see also movie S3). Note that we conducted rolling experiments using shadowgraphy under different magnifications (see `Imaging experiments and analysis', \emph{Supplementary Information} \cite{SI}). However, to facilitate subsequent image processing and analysis, we chose only those experiments performed at 50-60 frames per second acquisition rates and using a 4x magnification microscope objective lens providing a spatial resolution of 7-8\,$\mu$m/pixel \citep{SI}. 
To enhance observation, we conducted additional experiments by dispersing 100\,$\mu$m sized microplastics on the PDMS substrates along the rolling path. These microplastics adhered to the rolling PAAm sphere, acting as markers to help visualize the rolling motion (see movie S6). Note that the microplastics do not create or sustain the rolling motion in any form and the initial PDMS location where the PAAm sphere was placed contained no microplastics (see movie S6). 
We repeated the rolling experiments 15-20 times to ensure consistency \citep{SI}. 

Here, we analyze the experimental observations shown in Fig.~\ref{fig:1}.
For ease of analysis, without any loss of generality, 
we choose a window of the rolling event where the rolling sphere is completely visible and has traversed a vertical distance of approximately 1\,mm in 2 seconds. Consequently, we extract crucial dynamical features like evolution of the advancing edge $x_{\rm A}$, receding edge $x_{\rm R}$, center-of-mass $x_{\rm COM}$, and contact diameter $d$ (Fig.~\ref{fig:2}(a)). We also extract the contact angles at the advancing edge $\theta_{\rm A}$ and the receding edge $\theta_{\rm R}$. 
As the sphere rolls down, we observe an almost linear growth of the advancing and receding edges as well as the center-of-mass (Fig.~\ref{fig:2}(b)). 
Upon subsequent analysis, we observe that during the rolling motion, the average center-of-mass velocity $v_{\rm COM}$ and the average angular velocity $\omega$ are $0.51\,\rm{mm/s}$ and $0.94\,\rm{rad/s}$, respectively. Here, note that $\omega = 0.94\,{\rm{rad/s}} \approx 0.15\,{\rm s^{-1}}\neq \frac{v_{\rm COM}}{R} = 0.51\,{\rm s^{-1}} $. Thus, the observed motion is \emph{rolling with slipping} rather than \emph{pure rolling}. This feature becomes more evident from our rolling experiments with microplastics aiding in measuring $\omega$ with higher accuracy (Fig.~S10 \citep{SI}). 
Interestingly, we observe that the contact diameter is not fixed during the rolling sequence and continuously changes, oscillating about a mean value of $285\pm21\,{\mu}$m (Fig.~\ref{fig:2}(c)). Further, we observe that the dynamic contact angles at both the edges are greater than 160$^{\circ}$ with the contact angle at the advancing edge consistently larger than at the receding edge with an average hysteresis $\delta\theta = \theta_{\rm A} - \theta_{\rm R} = 2.9^{\circ}\pm1.6^{\circ}$ (Fig.~\ref{fig:2}(d)).
We note that the oscillation of the contact diameter indicates a continuous creation and destruction of the interface creating the torque and friction necessary to sustain this rolling motion. 

\begin{figure}[h!]
\begin{center}
\includegraphics[width=0.49\textwidth]{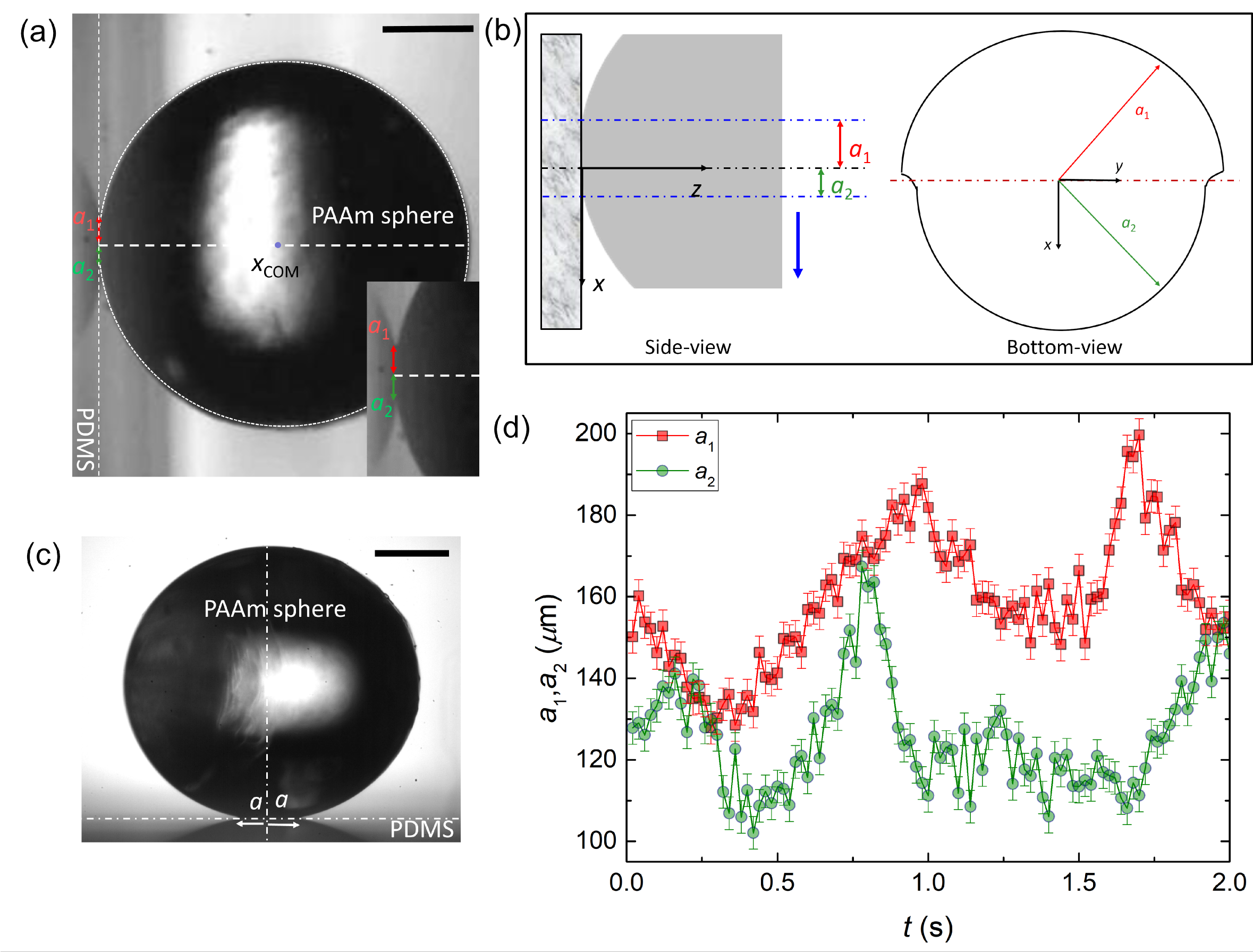}
\caption{(a) Experimental snapshot during rolling of a 1 mm radius PAAm sphere (elasticity, $E_1$ = 169.7 kPa) on a vertical PDMS substrate (elasticity, $E_{2}$ = 2422 kPa). The asymmetric contact interface is highlighted using the fitted circle (white dotted line) along the center-of-mass $x_{\rm COM}$ enabling extraction of the two unequal contact radii $a_{1}$ and $a_{2}$. Inset shows a close up of the contact interface. Scale bar represents 0.5\,mm. (b) Side-view and bottom-view schematic of the contact asymmetry. The blue arrow represents the direction of roll. (c) Static configuration of a similar 1 mm radius PAAm sphere (elasticity, $E_1$ = 169.7 kPa) on a \emph{horizontal} PDMS substrate (elasticity, $E_{2}$ = 2422 kPa). The symmetric contact interface is highlighted using equal contact radius $a$. Scale bar represents 0.5\,mm. (d) Evolution of $a_{1}$ and $a_{2}$ for the rolling of the PAAm elastic sphere ($E_{1} = 169.7\,{\rm kPa}$) on a vertical PDMS substrate ($E_{2} = 2242\,{\rm kPa}$), i.e., corresponding to the analysis in Fig.~\ref{fig:2}.}
\label{fig:3}
\end{center}
\end{figure}

Upon closer inspection, we noticed that the contact interface is not completely axisymmetric unlike its static counterpart on a horizontal substrate (Figs.~\ref{fig:3}(a)-(c)). By fitting a circle over the sphere profile in successive frames, we reveal the asymmetry in the contact diameter. This aids us in identifying the two unequal semi-circles with radii $a_{1}$ and $a_{2}$, where the subscripts 1 and 2 represents the contact radius in the receding and advancing edge, respectively and $a_{1}+a_{2} = d$ (Figs.~\ref{fig:3}(a),(b)). 
Consequently, we observe that $a_{1}$ and $a_{2}$ oscillate around 
mean values of $159\pm16\,{\mu}$m and $126\pm13\,{\mu}$m, respectively with $a_{1}$ consistently larger than $a_{2}$ (Fig.~\ref{fig:3}(d)). Using this observation of contact asymmetry, we invoke established crack theory and previous works by Dominik and Tielens \cite{dominik1995resistance}, Kendall \cite{kendall1975rolling}, and Krijit and co-workers \cite{krijt2014rolling} to interpret the rolling motion.

As the sphere rolls down, crack propagates through the contact interface, acting like an adhesive joint. Consequently, the crack at the advancing (leading) edge continuously closes whereas the crack at the receding (trailing) edge continuously opens resulting in the observed contact asymmetry. A rough estimate of this crack opening and closing velocity can be obtained from evaluating $dx_{\rm R}/dt$ (receding edge) and $dx_{\rm A}/dt$ (advancing edge) whose magnitudes averaged over the rolling duration are 0.54\,mm/s and 0.53\,mm/s, respectively (see also Fig.~S11 \citep{SI}). Interestingly, these velocities are approximately the same as the linear velocity of the center-of-mass, i.e., $v_{\rm COM} = 0.51\,{\rm mm/s}$. Higher magnification visualization of the rolling event confirms this crack opening/closing phenomenon (see movie S7).  

The strain energy release rate associated with the crack propagation process can be expressed as, $G_{1,2} = \frac{E^{*}}{2\pi}\frac{({a_{1,2}}^2-R{\delta})^{2}}{{a_{1,2}R^{2}}}$, where $G_{1}$ and $G_{2}$ are the energy release rates at the receding (crack opening) and advancing (crack closing) edges, respectively, and $G_{1} - G_{2}$ represents the \emph{adhesion hysteresis} \cite{kendall1975rolling}. Here, $\delta$ is the common indentation depth shared by both contact radii, $a_{1}$ and $a_{2}$. 
A natural consequence of the adhesion hysteresis is a finite torque and as a result, a finite friction force. In other words, rolling can only occur if $G_{1} - G_{2} > 0$, i.e., the energy required for opening the crack is higher than that required to close it. 
Using the contact asymmetry and the differential strain energy release rates, Dominik \emph{et al.} exhibited how the pressure distribution varies along the contact interface generating a torque about the y-axis (Fig.~\ref{fig:3}(b)), most pronounced at the periphery of the receding part of the contact interface \cite{dominik1995resistance}. 
This torque which essentially counters $mgR$ can be expressed as, $\tau = C_{0} R(a_{1} G_{1} - a_{2} G_{2})$ (see detailed derivation in \emph{Supplementary Information} \citep{SI} and also Ref.~\cite{krijt2014rolling}). Note that in the expression of torque, the prefactor $C_{0}$ can vary between 0.5 to 2 depending on the geometry of the contacting pairs. For sphere-sphere contacts, $C_{0} = \pi/2$. In the present study, we consider the mean of these limits, i.e., $C_{0} = 1.25$. Consequently, we can express the friction force $f$ as $f = C_{0}(a_{1} G_{1} - a_{2} G_{2})$. Thus, the force balance reads $ma_{\rm COM} = mg - f$ where $a_{\rm COM}$ is the linear acceleration of the center-of-mass.

Before proceeding with further analysis, here we make certain considerations relevant to the problem. 
First, typically the indentation depth $\delta$ is significant for static, horizontal contacts \cite{hertz1882beruhrung,johnson1971surface}. However, for the present vertical contact configuration and subsequent rolling, it is negligible due to the continuous rolling motion providing negligible time for the indentation to form. Thus, we assume ${a_{1,2}}^{2} >> R{\delta}$. Further, so far, we have considered the static Young's modulus for describing the elastic properties of the contacting pairs \cite{SI}. However, during rolling, the PAAm sphere induces contact on the underlying PDMS substrate at a finite angular velocity $\omega = 0.94\,{\rm rad/s} \approx 0.15\,{\rm Hz}$. Thus, we calculate the effective elasticity $E^{*}$ as  $E^{*}\vert_{\omega = 0.15\,{\rm Hz}} = [(1-{\nu_{1}}^2)/E_{1}\vert_{\omega = 0.15\,{\rm Hz}} + (1-{\nu_{2}}^2)/E_{2}\vert_{\omega = 0.15\,{\rm Hz}}]^{-1}$. 
As a result, for the specific combination of PAAm elastic sphere and PDMS substrate which exhibits the rolling motion, we have $E_{1}\vert_{\omega = 0.15\,{\rm Hz}} = 405\,{\rm kPa}$  and $E_{2}\vert_{\omega = 0.15\,{\rm Hz}} = 2316\,{\rm kPa}$ providing $E^{*}\vert_{\omega = 0.15\,{\rm Hz}} = 457\,{\rm kPa}$ (see also Fig.~S12 \citep{SI}).

\begin{figure}[]
\begin{center}
\includegraphics[width=0.49\textwidth]{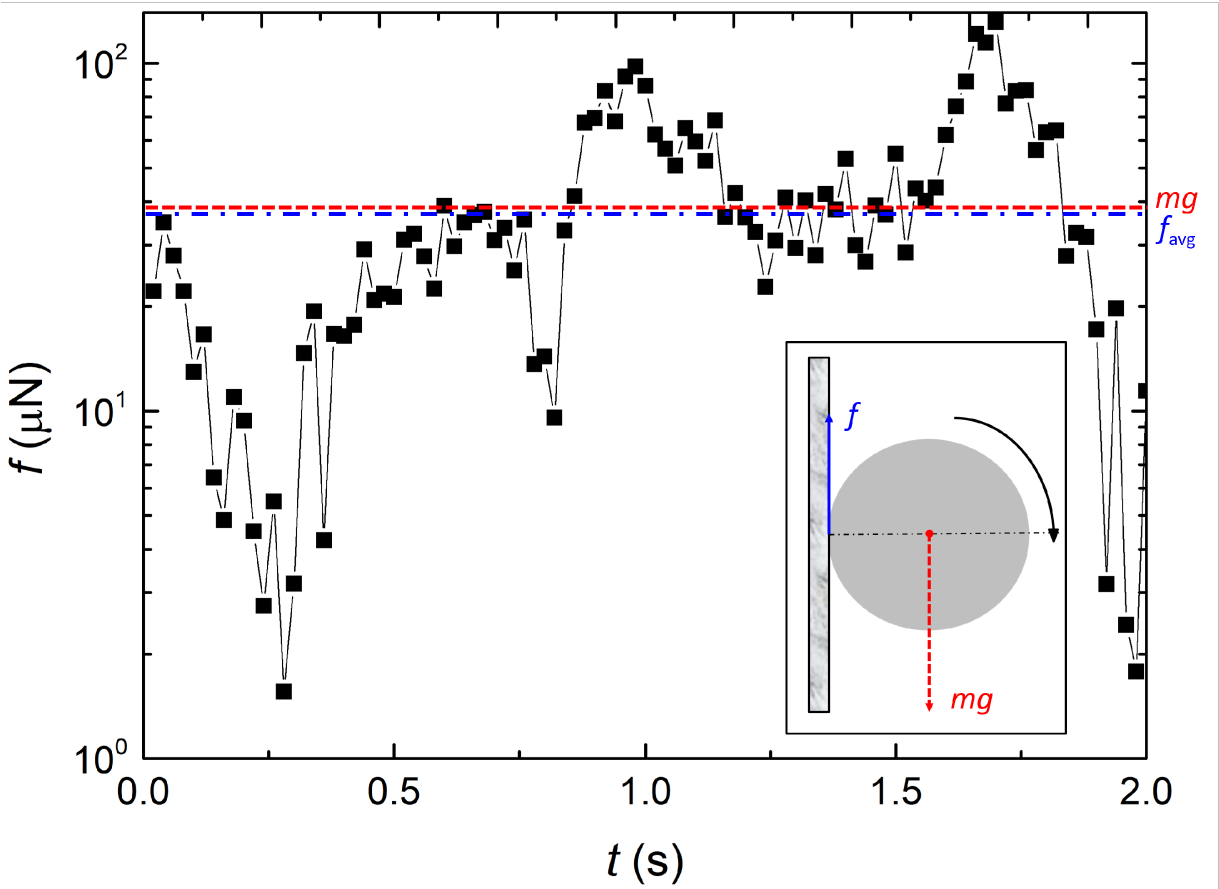}
\caption{Evolution of computed friction force $f$ for the rolling of a PAAm  sphere ($E_{1} = 169.7\,{\rm kPa}$) on a vertical PDMS substrate ($E_{2} = 2242\,{\rm kPa}$). $f_{\rm avg}$ and $mg$ are the average friction force and weight of the PAAm sphere, respectively. Inset shows the free body diagram of the process.}
\label{fig:4}
\end{center}
\end{figure}
In Fig.~\ref{fig:4}, we reveal the variation of the experimentally calculated friction force $f$ over the duration of rolling event. Since $f$ is a function of $a_{1}$ and $a_{2}$, it also oscillates about a mean value $f_{\rm avg} = 38.7\,{\mu}{\rm N}$ during the course of rolling. Note that since $v_{\rm COM}\sim 10^{-4}\,{\rm m/s}$ and $a_{\rm COM}\sim 10^{-4}\,{\rm m/s^{2}}$, $ma_{\rm COM} << mg$ Thus, from the force balance expression, it is expected that $f \approx mg$ which is reflected in our experiments where $f_{\rm avg} = 38.7\,{\mu}{\rm N} \approx mg = 39.2\,\mu{\rm N}$. 
Thus, for rolling on a 90$^{\circ}$ incline, the average friction force is just enough to balance the weight of the rolling sphere. 
Conversely, the torque expression reads $mgR - fR = I\alpha$ and is perfectly balanced since $I\alpha << mgR, fR$. Here $I = (2/5)mR^{2}+(1/2)mR^{2}$ is the moment of inertia, $\alpha$ is the angular acceleration, and $I\alpha \sim 10^{-6}\,\mu{\rm N.m}$ whereas $mgR, fR \sim 10^{-2}\,\mu{\rm N.m}$. 


Lastly, we perform a conservation of energy analysis to interpret this rolling motion. To do so, we consider two instances at the beginning and end of the chosen window of the rolling event shown in Fig.~\ref{fig:1} and analyzed subsequently in Figs.~\ref{fig:2}-\ref{fig:4}. The energy balance dictates $\frac{1}{2}m{v_{\rm i,COM}}^{2}$ + $\frac{1}{2}I{\omega_{\rm i,COM}}^{2} + mg\Delta h = \frac{1}{2}m{v_{\rm f,COM}}^{2}$ + $\frac{1}{2}I{\omega_{\rm f,COM}}^{2} + f_{\rm avg}\Delta h  $ where $I$ is the moment of inertia and $\Delta h \approx 1\,{\rm mm}$ is the vertical distance traversed by the rolling sphere in the chosen window.  Here, the subscripts i and f represent the initial and final states, respectively. Upon performing an order of magnitude analysis, we observe that $\frac{1}{2}I{\omega_{\rm COM}}^{2} \sim \mathcal{O}({10^{-9}\,\mu{\rm N.m}}) << mg\Delta h$, $f_{\rm avg}\Delta h \sim \mathcal{O}({10^{-2}\,\mu{\rm N.m}}) $ and $\frac{1}{2}m{v_{\rm COM}}^{2} \sim \mathcal{O}({10^{-6}\,\mu{\rm N.m}}) << mg\Delta h, f_{\rm avg}\Delta h \sim \mathcal{O}({10^{-2}\,\mu{\rm N.m}}) $. Thus, the energy balance simplifies to $mg\Delta h = f_{\rm avg} \Delta h$ which aligns with our previous force balance analysis, i.e., $mg \approx f_{\rm avg}$ (Fig.~\ref{fig:4}). In essence, the observed rolling motion satisfies conservation of energy, where the potential energy is entirely dissipated by the work done by the kinetic friction. At the same time, our energy analysis theoretically indicates that the rolling motion will continue as long as there is a finite $\Delta h$ inducing finite, non-zero potential energy. This is confirmed in our experiments, where we observed rolling events terminating at $\Delta h = 4\,{\rm mm}$ (see movie S8) as well as events where the sphere rolled down with $\Delta h> 15\,{\rm mm}$, exceeding the substrate's length eventually (see movie S9). Therefore, for rolling to terminate, additional dissipation mechanisms due to physical and chemical heterogeneities at the contact interface are essential.

In the preceding sections, we analyzed the conditions that sustain the rolling motion.
Here, we briefly discuss the probable reasons for the onset of rolling. Initially, after the PAAm sphere is deposited, $G_{1} < G_{\rm op}$ where $G_{\rm op}$ is the critical strain energy required to open the crack at the trailing edge \cite{dominik1995resistance,kendall1975rolling}. Rolling begins only when $G_{1}$ overcomes $G_{\rm op}$, causing the crack to open at the trailing edge and leading to the adhesion hysteresis necessary to initiate rolling. Factors contributing to this include mechanical events like high strain rates at the trailing edge or chemical events such as molecular rearrangement at the contact interface \cite{chen1991molecular}. 
Conversely, a similar condition can also prematurely terminate the rolling; if $G_{1}$ instantaneously drops below $G_{\rm op}$, the crack at the trailing edge fails to open, stopping the roll. 
However, a detailed understanding of the conditions that dictate the onset or termination of rolling remains a topic of future work.

In conclusion, our theoretical and experimental findings highlight the conditions under which rolling can occur on a 90$^{\circ}$ incline. The primary requirement is adhesion hysteresis at a soft-soft interface, which continuously opens at one end and closes at the other, similar to a crack. This pressure asymmetry generates the torque and friction force necessary to sustain continuous rolling motion. Our findings can guide the design of soft micro rovers for navigating unpredictable terrain during space exploration. Additionally, understanding this type of rolling motion is relevant for N-body simulations of dust aggregate compression in protoplanetary discs \cite{kataoka2013static}. Such aggregates consist of spherical monomers whose rolling-driven mutual interactions are crucial for understanding compression \cite{kataoka2013static,dominik1997physics}. 

\emph{Acknowledgment}: The authors thank Prof. Michael
K.C. Tam (Department of Chemical Engineering, University of Waterloo) providing the dynamic shear rheometer. B.Z. acknowledges the support of NSERC
RGPIN-2019-04650. S.K.M acknowledges the support of the Discovery Grant (NSERC, RGPIN-2024-03729).
Finally, the authors acknowledge the use of Microsoft's Copilot (version GPT-4) for paraphrasing and proof reading parts of the manuscript.


\bibliography{ref}

\end{document}